\documentstyle[11pt,epsf]{article}
\newcommand{\beq}{\begin{equation}}
\newcommand{\eeq}{\end{equation}} 

\newcommand{\beqa}{\begin{eqnarray}}
\newcommand{\eeqa}{\end{eqnarray}}

\def\opone{\leavevmode\hbox{\small1\kern-3.8pt\normalsize1}}

\begin{document}

\title{Twin-photon techniques for fiber measurements}
\author
{N. Gisin J. Brendel and H. Zbinden\\
\protect\small\em Group of Applied Physics, University of Geneva, 
1211 Geneva 4, Switzerland \\
A. Sergienko and A. Muller \\
\protect\small\em Dept. of Electrical and Computer Engineering, Boston University}

\maketitle

\section{Introduction}\label{introduction}
%=========================================================================
Characterization of single-mode optical fibers requires a basis for time, wavelength and
polarization. Parametric downconversion in non-linear
crystals naturally provide pairs of photons extremely highly correlated in time,
energy (thus wavelength) and polarization. It is thus tempting to explore such photon pairs
for fiber and fiber device characterization. Historically, entangled photon pairs where first used
in delicate tests of quantum mechanics \cite{testQM}. Indeed, their correlation is higher than
classically possible! 
However, nowaday photon pair sources can be
made cheap and compact enough to offer practical alternatives for many of the
traditional measurement schemes. In addition photon pairs provide entirely new possibilities
and open the door to new developments in metrology. The purpose of this contribution is
first to present some concrete proposals and results along the above lines,
and next to draw the attention of the audience to these rather revolutionary new
combinations of quantum optics and communication.

\section{A photon pair source}\label{sectBell}
%======================================================================

\begin{figure}
%fig1
\begin{center}
\leavevmode
\epsfxsize=60mm
\epsffile{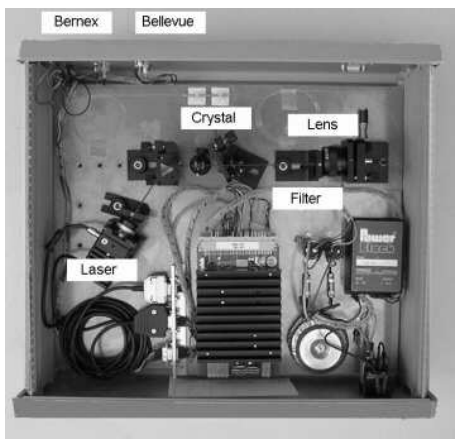}  
\caption{Example of a compact twin-photon source.}
\end{center}
\end{figure}

Figure 1 presents an example of a photon pair source. A non-linear crystal, like
BBO, $KNbO_3$, $LiIO_3$, is pumped by a 15 mW laser diode, at 660 nm in the present example. 
The prism between the laser diode and the crystal removes all extra infra-red light.
The crystals are well known for frequency doubling, as in doubled Nd-YAG lasers. 
Here we use the time reversed process:
wavelength doubling, better known as downconversion 
(recall that all elementary quantum processes are time symmetric).
Since initially the infrared modes are empty (ie are in their quantum vacuum state
and the power of pump radiation is not very strong), this
downconversion process is spontaneous: a photon from the pump spontaneously splits into
two infrared photons. Energy and momentum conservation forces these two photons to be 
highly correlated: they are created exactly at the same time and the sum of their energy
(optical frequency) equals the energy of the pump photon. Note that the latter is defined
within the spectral width of the laser, typically expressed in MHz, hence the sum of the
two downconverted photon is well defined. But the spectral width of each 
downconverted photon
can be quite large, up to tens of nm, or even hundreds of nm. Following the optical path
on Fig. 1, the photon pairs are then injected into a single mode fiber, with standard techniques
(microscope objective in the shown example). Next, a directional coupler separates both 
photons and guides them to two optical connectors. 

Figure 1 presents one possible implementation. Other examples are discussed in the next sections.
For instance, it might be useful to replace the directional coupler by a wavelength
demultiplexer: the two photon are still quantum correlated (ie extremely highly correlated)
in time and energy, but the photon at one output is in the short wavelength end of the 
spectrum, while the companion photon is in the complementary long wavelength range.
In fig. 1 a collinear configuration is present: the pump and both downconverted photons
follow parallel trajectories. Another possibility exploits momentum conservation: if one
photon emerges at an angle with respect to the pump beam, the companion photon emerges
on the opposite angle (the detail depend on the exact phase matching conditions,
wavelength filters, etc). This will be illustrated in the next section for various
dispersion measurements.

So far we did not mention polarization, but since the crystals are anisotropic, polarization
is important (actually birefringence is required for the phase matching condition). 
There are two kinds of downconversion processes. In type I both
infrared photons have identical polarization, both orthogonal to the pump laser's
polarization. In type II downconversion, on the contrary, the two infrared photons can be
totally unpolarized, however their polarization states are quantum correlated: if one
passes through a polarizer, the other one is immediately polarized in the 
orthogonal direction. This will be illustrated in subsection \ref{PMD} for
polarization mode dispersion measurement.

Other examples of interesting configurations for 2-photon sources are described in
\cite{InnsbruckSource,IQEC-Kwiat}.

\section{Photon counters}
%========================
Photomultipliers exist for years, but are not effective at telecom wavelengths (1.3 and 1.55
$\mu$). Photon counting devices based on silicon Avalanche Photo-Diodes (APD) are 
commercially available, but also limited to wavelengths below 1 $\mu$. However, a few groups
around the world have demonstrated that germanium and InGaAs APD are suitable for photon
counting at telecom wavelengths \cite{PhotonCount}. Admittedly, this is still
the weak point of "quantum optics at telecom wavelengths", because most of the demonstrations
used liquid nitrogen cooled APD. Actually, very recently we have demonstrated that
InGaAs APDs Peltier cooled are suitable for metrology applications 
where noise requirement is not as severe as for quantum communication applications. 
Anyway, progress in this field is fast. The fact that until
now not much efforts have been put into these developments, allows one to be optimistic
about a possible breakthrough in the near future.

\section{Dispersion measurements}
%===============================================
The intrinsic time-energy-polarization correlation of the downconverted
photons make them natural candidates for
dispersion measurements, as illustrated in the next 3 subsections.

\subsection{Chromatic dispersion measurements on long fibers}
%================================================
This is possibly the simplest example. It exploits only the time and energy correlations
of the photons: the difference in detection times provides the information on the time of
flight and the measurement of the wavelength of one photon is enough to obtain the
necessary information on the wavelength of the companion photon. Therefore we have direct 
access to the group delay between complementary wavelengths. Choosing moreover the central
wavelength near the zero chromatic dispersion, these delays will be small, thus not
necessitating any long electronic delay lines. In short, this technique makes use of the
photon pairs as a broadband lightsource, comparable to a LED, with sub-picosecond
pulsewidth.

\begin{figure}
%fig2
\begin{center}
\leavevmode
\epsfxsize=120mm
\epsffile{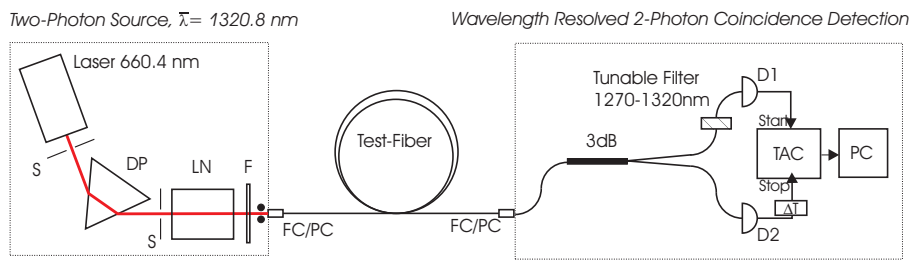}  
\caption{Setup for chromatic dispersion measurements.}
\end{center}
\end{figure}

A possible implementation is presented in Fig. 2. For technical details we refer the interested
reader to \cite{Brendel98}. In our first experiment we could determine the zero chromatic
dispersion wavelength with a precision of $\pm 0.5$ nm \cite{Brendel98}. In a forthcoming
experiment, we expect to achieve higher resolution, larger spectral width (up to 300 nm) and
measurement times of minutes. 

\subsection{Chromatic dispersion measurements on short fibers}
%================================================
\begin{figure}
%fig3
\begin{center}
\leavevmode
\epsfxsize=100mm
\epsffile{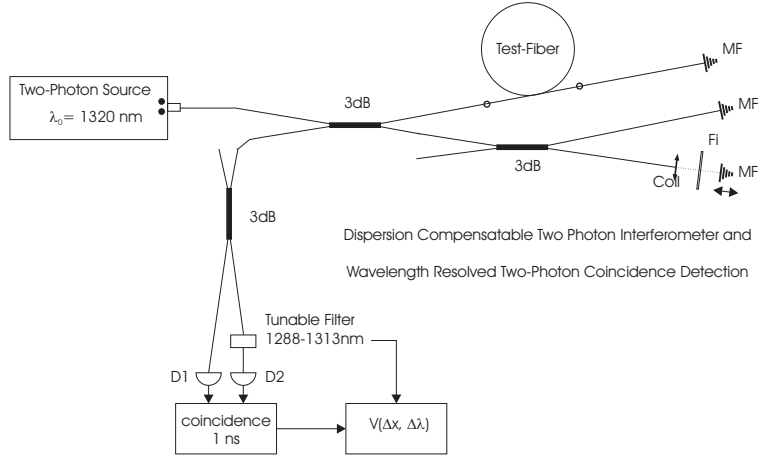}  
\caption{Setup 2-photon interferometric measurements.}
\end{center}
\end{figure}

The spectra of the downconverted photons are large, like LEDs or even broader. However, 
the photon pairs, considered as a whole, still enjoy the high coherence of the pump laser.
Hence, using 2-photon interferometry, the chromatic dispersion of meters long samples of
fiber can be determined. This is similar to the well-known white light interferometric
measurement technique \cite{WhiteLightInterf}, but with the advantage that the fiber sample length
does not need to be adjusted with millimeters accuracy. Thanks to the meter long coherence
of the pump laser, the sample length needs only to be adjusted within cm or meter precision.
The proposed setup is presented in Fig. 3. Admittedly, this example is more complex than
the previous one. Depending on the temperature of the crystal the wavelengths of the 
collinearly emitted photons vary between 1250 and 1350 nm for one photon and between 
1450 and 1600 nm for the other, the central wavelength being at 1400 nm. The 
interferometer is unbalanced, having the test-fiber in the long arm and a wavelength 
division multiplexer (WDM) and a moveable mirror in the short arm. The two photon 
have four choices: The first one can either take the long arm and the other one the short 
one or vice versa, or both photons take either the short or the long arm. The first two 
cases are distinguishable due to the time difference between the arrival times of the two 
photons. The latter two possibilities are indistinguishable, hence lead to two photon 
interference. The visibility of the interference pattern is reduced, if the dispersion is not 
identical in the two arms of the interferometer. We can compensate for the difference in 
the group velocities in test fiber between the photon around 1300 nm and the photon 
around 1500 by separating them with the WDM in the short arm and by introducing an 
adjustable delay for one of them by a moveable mirror. The position of this mirror for 
optimal interference visibility is a measure for the time delay between the two 
wavelengths introduced by the dispersion in the test fiber. We can now scan over the 
whole wavelength range by tuning the temperature of the nonlinear crystal and obtain the 
dispersion between 1250 and 1600 nm, thus obtain the group delay as function of wavelength.

\subsection{Polarization mode dispersion measurements}\label{PMD}
%================================================
\begin{figure}
%fig4
\begin{center}
\leavevmode
\epsfxsize=90mm
\epsffile[40 340 542 620]{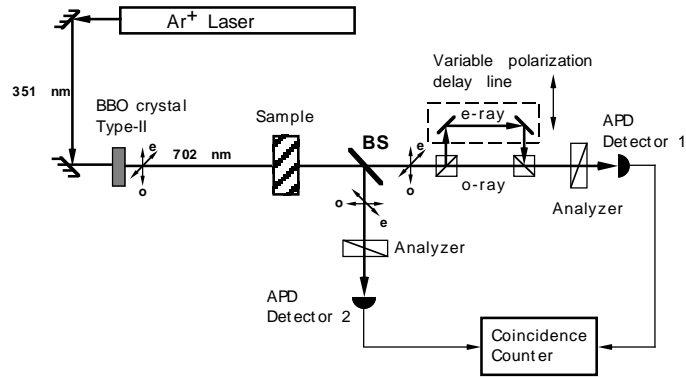}  
\caption{Setup for PMD mesurements.}
\end{center}
\end{figure}

In this example, the photon pairs are produced by type II parametric downconversion. Hence,
in addition to their time and energy correlations, they are also entangled in polarization,
thus suitable for Polarization Mode Dispersion (PMD) measurements. A possible
implementation is presented in Fig. 4, for details see ref. \cite{SergienkoPMD}. In this
experiment, birefringent plates where actually measured, not fibers, and the source was
an Argon laser. However, the results clearly demonstrate the potential of the method,
see Fig. 5. 
Essentially the method is similar to the white light interferometric technique, but thanks
to the wide spectrum of the downconverted photons (about 300 nm is this experiment),
a resolution of 0.01 fs was achieved. This should be contrasted with the 50 to 100 fs resolution
of the standard interferometric technique based on LEDs \cite{PMDCOST241}. 

\begin{figure}
%fig5
\begin{center}
\leavevmode
\epsfxsize=120mm
\epsffile[118 384 611 542]{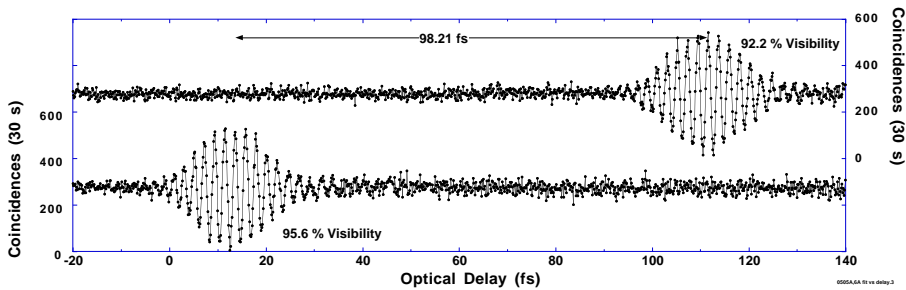}  
\caption{Result of the PMD measurement with sub-femtosecond resolution.}
\end{center}                                                           
\end{figure}

\section{Absolute calibration of sources and detectors without black-body radiation}\label{secmutinfo}
%====================================================================================
As a last example, we present the use of photon pairs for the determination
of the absolute efficiency of detectors. Let us emphasize that this technique, contrary to all
other techniques we are aware of, is not based on a light source calibrated with
black body radiation. Hence, this new technique 
allows to determine detector efficiencies independently
of temperature. The basic idea is simple \cite{DetectEff}. Since the downconverted photons are always
produced by pairs, the number $N_{AB}$ of coincidence detections with two detector A and B equals:
\beq
N_{AB}=N\eta_A\eta_B
\eeq
where $N$ is the total number of photon pairs, and $\eta_A$ and $\eta_B$ are the efficiencies
of detectors A and B, respectively. Moreover the single counts at detector B equals
$N_B=N\eta_B$. Consequently, the efficiency of detector A is given by the simply relation:
\beq
\eta_A=\frac{N_{AB}}{N_B}
\eeq
This elegant idea has now been tested in several labs \cite{DetectEff}, in particular at 
the NIST-Washington, and the accuracy limitations due
to the statistics of photon counting has been estimated \cite{DetectEffStat}. A trial to extend this
technique to analog detectors is presented in \cite{AnalogDetectEff}.

Analogoulsy, a calibrated source can be devised.

\section{Conclusion}\label{concl}
%================================
The potential of highly correlated photon pairs produced in parametric downconversion 
for optical fiber and devices has been emphasized. The huge technological progress in simple
and efficient photon pair sources and the ongoing progress in photon counting techniques
at telecom wavelengths open new possibilities for the fiber optic community. The intrinsic
correlations of the photon pairs offer advantages such as automatic determination of
a common time basis, complementary wavelengths and polarization. Thus all the degrees of
freedom of photons in single-mode fibers are entangled in these photon pairs. Each photon
may be totally depolarized and cover a bandwidth of tens or even hundreds of nanometers. Now,
at the same time the two photons form a single quantum object (a bi-photon) with large
coherence lengths, up to meters. This offers the possibility to combine advantages
of low and high coherence interferometry.

The examples presented in this contribution are but a few ones, more should be expected.
Finally, let us briefly mention that quantum optics has also interesting 
potentials for communication. As examples we mention quantum cryptography \cite{PhysWorldQC}
and noiseless amplifiers that beat the 3dB "quantum limit"\cite{Qamplifier}.

\section*{Acknowledgments}
%=========================
This work profited from support by the Swiss Priority Program in Optics and by the 
European TMR network on "The physics of quantum information".

%\newpage


\small
\begin{thebibliography}{99}
%===========================

\bibitem{testQM} see e.g. J. Freedman, and J. F. Clauser, Phys. Rev. Lett., {\bf28},938-941, (1972); 
             A. Aspect, P. Grangier, and G. Roger, Phys. Rev. Lett., {\bf47}, 460-463, (1981); 
             Z. Y. Ou and L. Mandel, Phys. Rev. Lett., {\bf61}, 50-53, (1988); 
             Shih and Alley, Phys. Rev. Lett. {\bf61}, 2921, (1988);
             P. R. Tapster, J. G. Rarity, and P. C. M. Owens, Phys. Rev. Lett., 
                 {\bf73}, 1923-1926, (1994). 
             W. Tittel et al., Phys. Rev. A {\bf57}, 3229, 1998.

\bibitem{InnsbruckSource} P. G. Kwiat et al., Phys. Rev. Lett., {\bf75}, 4337, (1995).   ??? nb400


\bibitem{IQEC-Kwiat} P. Kwiat, CLEO/IQEC, pp, San-Fransisco, 1998.

\bibitem{PhotonCount} F. Zappa et al, Optics Lett. {\bf 19},846-848,1994;
           P.C.M. Owens et al, Applied Optics {\bf 33},6895-01,1994;
           G.L. Morgan et al, Los Alamos National Laboratoty report LA-UR-97-4375, 1997;
           G. Ribordy et al, Applied Optics, in press, 1998.

\bibitem{Brendel98} J. Brendel, H. Zbinden and N. Gisin, Optics Commun., in press, 1998.

\bibitem{WhiteLightInterf} L. Thévenaz et al, J. Lightwave Tech. {\bf 6}, 1-7, 1988.

\bibitem{SergienkoPMD} A.V. Sergienko et al., preprint, Boston University, 1998.  

\bibitem{PMDCOST241} N. Gisin et al, JEOS Pure \& Applied Optics {\bf4}, 511, 1995.

\bibitem{DetectEff} see e.g. A.V.Sergienko and A. N. Penin, Applied Optics, {\bf 30}, 3582, 1991;
           A.L. Migdall et al., Metrologia {\bf32}, 479-483, 1995;
           G. Brida et al, preprint, Istituto Electtrotecnico Nazionale G. Ferraris, 
           Torino, Italy, 1998.

\bibitem{DetectEffStat} P. Kwait et al, Applied Optics {\bf33}, 1844, 1994; 
           N. Gisin, J. Brendel and A. Stefanov, preprint, Geneva University, 1998.

\bibitem{AnalogDetectEff} A. Sergienko and A.N. Penin, Sov. Tech. Phys. Lett. {\bf12}, 328, 1986.

\bibitem{PhysWorldQC} W. Tittel, G. Ribordy and N. Gisin, Phys. World, March 1998.

\bibitem{Qamplifier} D.J. Lovering et al, Optics Lett. {\bf21}, 1439, 1996.



\end{thebibliography}
\end{document}